# Geometry-dependent skin effects in reciprocal photonic crystals

Zhening Fang, Mengying Hu, Lei Zhou*, and Kun Ding[†]

Department of Physics, State Key Laboratory of Surface Physics, and Key Laboratory of Micro and Nano Photonic Structures (Ministry of Education), Fudan University, Shanghai 200438, China

*Email: phzhou@fudan.edu.cn

[†]Email: kunding@fudan.edu.cn

**Abstract:** Skin effect that all eigenmodes within a frequency range become edge states is dictated by the topological properties of *complex* eigenvalues unique in *non-Hermitian* systems. The prevailing attempts to realize such a fascinating effect are confined to either one-dimensional or nonreciprocal systems exhibiting asymmetric couplings. Here, inspired by a recent model Hamiltonian theory, we propose a realistic reciprocal two-dimensional (2D) photonic crystal (PhC) system that shows the desired skin effect. Specifically, we establish a routine for designing such non-Hermitian systems via revealing the inherent connections between the non-trivial eigenvalue topology of order-2 exceptional points (EPs) and the skin effects. Guided by the proposed strategy, we successfully design a 2D PhC that possesses the EPs with nonzero eigenvalue winding numbers. The spectral area along a specific wavevector direction is then formed by leveraging the symmetry of the macroscopic geometry and the unit cell. The projected-band-structure calculations are performed to demonstrate that the desired skin effect exists at the specific crystalline interfaces. We finally employ time-domain simulations to vividly illustrate this phenomenon by exciting a pulse at the center of a finite-sized PhC. Our results form a solid basis for further experimental confirmations and applications of the skin effect.

# 1 Introduction

Band topology has become a key ingredient in physics because it conveys the profound connotation of the eigenwavefunction of a system exhibiting certain global symmetry [1]. Bulk-edge correspondence endows the robustness of topological edge modes, and a vast myriad of interesting phenomena has then been realized [1-5]. In a Hermitian system, topological invariant roots in a bundle of eigenvectors (wavefunctions) attained by continuously varying the parameters involved in the Hamiltonian [6-10], while all eigenvalues do not carry any topological information but only tell the gap positions [2]. Recent development in *non-Hermitian* theory, however, unveils the eigenvalues can also exhibit topological signatures due that the eigenvalues for a non-Hermitian Hamiltonian naturally live in the complex plane [11]. Such a seemingly straightforward extension makes the eigenvalue winding numbers meaningful, adding another layer to the band topology, which is now called the spectral topology [12-16]. The emerging feature herein is that the band gaps in the complex plane can be classified into two types: the line gap and the point gap [17]. The line gap, literally speaking, is a gap where different energy bands are separated by a line and can be evolved into a traditional band gap through Hermitian flattening. However, the point gap is a singular point surrounded by the energy bands, which is unique in the non-Hermitian system. Since the bands surrounding the singularity always form a closed loop, the area formed by the loop is called a spectral area [17]. Such exciting concepts and the fact that an ideal Hermitian system is usually difficult to realize in real life have made non-Hermitian physics a vibrant field in the past few years [18-25].

Non-Hermitian skin effect (NHSE) is perhaps the most significant phenomenon coming from the nontrivial winding numbers of eigenvalues in the complex frequency (energy) plane [11, 26-54]. Different from the topological edge modes in a Hermitian system where a few edge states are found within the gap between bulk states, in a non-Hermitian system exhibiting the skin effect, *all* bulk states calculated under the periodic boundary condition (PBC) can collapse to structure edges of the system with the open boundary condition (OBC) imposed [30]. The NHSE was firstly predicted by S. Yao and Z. Wang in studying a one-dimensional (1D) model system with asymmetric inter-unit couplings [11] and was soon experimentally realized in a 1D photonic system [45-48], inspiring further research in the higher dimensional systems [54-68]. Very recently, C. Fang *et al.* predicted that the NHSE can also exist in a two-dimensional (2D) non-Hermitian system, as long as the PBC eigenvalues of the system for the wavenumber ***k*** along a particular line in the Brillouin zone (BZ) form a spectral area in the complex plane [54]. An intuitive physics picture is that, as such a condition is satisfied, (complex) eigenfrequencies of a particular ***k*** mode and its specular reflection mode with respect to the crystalline interface must be different, indicating that no standing waves (i.e., bulk modes) can be formed inside the system at these frequencies [17, 54]. In reciprocal systems, such a criterion can only be satisfied on specific paths in the BZ, and thus the desired skin effect can only be found at specific crystalline interfaces of the systems under study, which is now termed the geometry-dependent-skin-effect (GDSE) [54]. From the symmetry point of view, the GDSE obviously requires a mismatch between the geometric symmetry and the lattice symmetry [53, 54].

In this paper, we establish a routine for designing realistic reciprocal photonic systems exhibiting spectral topologies and numerically demonstrate the predicted GDSE in a particular system via full-wave time-domain simulations. Recognizing the intrinsic connections between the nonzero eigenvalue winding numbers of the exceptional points (EPs) and the spectral area, we utilize two pairs of order-2 EPs in a 2D PhC and leverage the symmetry to realize the spectral area and point gaps formed by the complex eigenfrequencies along a non-trivial path in the BZ, which fulfills the criterion of GDSE. We then perform the projected band structure calculations for the corresponding crystalline interfaces with the PBC (OBC) applied along (perpendicular to) the interface, and the spatial distributions of all eigenstates within the spectral range of interest clearly show the skin effect. We finally numerically study the time-evolutions of a pulse excited at the center of the photonic structures with trivial and non-trivial interfaces, vividly demonstrating the existence of the predicted GDSE at the non-trivial interfaces of the system.

**2 Spectral area and point gaps from stable exceptional points**

We start from designing a 2D reciprocal PhC exhibiting several pairs of stable order-2 EPs. The PhC consists of a square lattice of circular air holes embedded in dielectric, with unit cell shown in the top panel of Figure 1(a). Due to the lattice symmetry, there exists a two-fold degeneracy at the Γ point in the band structure. The bottom panel of Figure 1(a) shows the two bands forming the degeneracy in the $k_x - k_y$ plane calculated by a model Hamiltonian (see Section I in Supplemental Materials for details). A quadratic point (QP) is clearly seen at the Γ point and further confirmed by full wave simulations via the finite element method (FEM), which is shown in Figure 1(d). Solid and dashed lines herein respectively show the bands along the $\Gamma - Y$ direction calculated by the FEM and the model Hamiltonian. By deforming the square lattice into an orthorhombic lattice depicted in Figure 1(b), the QP gives rise to two Dirac points (DPs) in the $\Gamma - Y$ direction which are highlighted by the filled black circles in the left panel of Figure 1(b). From the group symmetry point of view, the Γ point belongs the $C_{4v}$ ($C_{2v}$) point group in the square (orthorhombic) lattice case, and thus the two-fold degeneracy protected by the $C_{4v}$ symmetry has been lifted by distorting the unit cell. However, two bands along the $\Gamma - Y$ direction still exhibit orthogonal symmetries, and hence DPs could be formed. The full band structure calculated by FEM is shown in the right panel of Figure 1(d) and marked by three different colors. The bands in red (black) color are lower (higher) frequency bands, which do not form the DPs and are not focused on in this work. The blue color highlights the bands forming the DPs, and the comparison between the FEM results and the model Hamiltonian is shown in the left panel. Good agreement between two results confirms the existence of Dirac cones in the orthorhombic lattice.

Once two DPs have been realized, two pairs of EPs are expected to be found in the 2D BZ when non-Hermiticity is present [69-71]. This is because two degrees of freedom (DOFs) are required to stabilize an order-2 EP with nonzero discriminant numbers (DNs) [69]. We assume that the refraction index of the dielectric medium is $n_D + i\gamma$ with $\gamma$ denoting the dissipative loss. In the case of $\gamma = 0.1$ which can be realized experimentally, we find that two DPs in Figure 1(b) give birth to four EPs nearby as shown in Figure 1(c). FEM-calculated values of Re(ω) [Im(ω)] for *k* points on the gray surface ($k_y a/2\pi = -0.075$) are plotted in Figure 1(f) by the blue [red] lines, from which two EPs marked by open black stars are clearly identified.

Note that the non-Hermiticity here does not have any explicit symmetry, and thus each DP only gives rise to a pair of EPs rather than an exceptional ring [70, 72]. To quantify the spectral winding near these EPs, we calculate the DN of each EP $v_{EP}$ by the equation [53]

$$v_{EP} = v_{\pm} + v_{\mp} \qquad v_{\pm(\mp)} = \frac{-1}{2\pi} \oint_{\Gamma_{EP}} d\boldsymbol{k} \cdot \nabla_{\boldsymbol{k}} \arg[\omega_{+(-)}(\boldsymbol{k}) - \omega_{-(+)}(\boldsymbol{k})], \qquad (1)$$

where $\Gamma_{EP}$ represents the infinitesimal counterclockwise loop in the BZ enclosing an EP with $\boldsymbol{k}$ being the relevant parameters, and $\omega_{\pm}(\boldsymbol{k})$ is the two eigenfrequencies forming the EP. Figure 2(a) shows the position of four EPs in the first BZ, and the EP in blue (red) indicates its DN is $+1$ ($-1$). For a loop encircling the EP, the nonzero DN implies that the eigenfrequencies enclose a spectral area, or equivalently speaking form a point gap, in the complex frequency plane, which is a basic ingredient in the NHSE [53]. As discussed previously, we shall find a set of parallel $\boldsymbol{k}$ loops along which the corresponding eigenfrequencies form a point gap, and then crystalline interfaces perpendicular to these $\boldsymbol{k}$ loops must support the GDSE [53]. Therefore, finding such non-trivial $\boldsymbol{k}$ loops is pivotal in realizing the desired GDSE.

We begin our consideration with the complete *k* loops in the first or extended BZ because otherwise, the eigenfrequency does not come back after travelling one circle. Take the route-1 labelled by the solid orange line in Figure 2(a) as the first example, which is obviously a complete loop. The FEM-calculated eigenfrequencies along it are plotted in the complex frequency plane by the solid orange lines in Figure 2(b). A line gap clearly separates the two bands forming the EP and thus no spectral areas are present, indicating the bands are spectral reciprocal along the horizontal direction. This is because the unit cell possesses the mirror symmetry along both *x* and *y* directions, and thus uncovers the role of symmetry here. The spectral area is expected to be formed in general, but matching the wavevector direction with the unit cell symmetry makes it vanish. Thus, these wavevector directions with zero spectral areas can be treated as trivial $\boldsymbol{k}$ loops, for instance, the vertical and horizontal directions here. Therefore, the non-trivial $\boldsymbol{k}$ loops shall be chosen not to possess any symmetry which guarantees spectral reciprocity. With this understanding in mind, we choose two parallel routes which do not exhibit any explicit symmetries, and label them as route-2 and route-3 as shown in Figure 2(a). The eigenfrequencies calculated by FEM along these two routes are shown by triangles in Figure 2(b) and circles in Figure 2(c), respectively. The spectral area spanned along each route is clearly nonzero, which is highlighted in green and blue. It is then expected that point gaps can be formed when the orientations of $k$ are parallel to the route-2 and route-3. Comparing with the blue region, the spectral regions in green have been separated into two portions. They are not connected, but the spectral area for each portion is still nonzero (see Section II in Supplemental Materials for details). The same conclusions can be drawn if we increase the non-Hermitian strength to $\gamma = 1$. As shown in Figure 2(d), the spectral area for the route-2 and route-3 is still nonzero but vanishes for the route-1. Hence, whatever the values of $\gamma$, we expect that GDSE appears at the crystalline interfaces perpendicular to route-2 and route-3 but does not at the interface dictated by route-1. We then consider all the $k$ loops other than the vertical and horizontal directions to complete the statement. From the previous argument, these $k$ loops shall be non-trivial, but the eigenfrequencies along these loops are not guaranteed to form a point gap solely. However, complementing several $k$ segments along the BZ boundary will help a point gap form, and the ensuing GDSE is also expected (see Section II in Supplemental Materials for details).

**3 Geometry dependent skin effects**

To verify the existence of skin effects for distinct crystalline interfaces, we perform projected band structure calculations to study the structure shown in the inset of Figures 3(a) and 3(c). The system consists of $N$ unit cells in the direction perpendicular to the crystalline interface of interest, and the boundary conditions (BCs) applied to both ends are perfect electric conductors (PECs). The BCs applied along the crystalline interface are periodic boundary conditions (PBCs), which simplify the considered problem into a 1D one and make the analysis traceable [51]. We denote the setup in Figure 3(a) [Figure 3(c)] as the horizontal (oblique) configuration which corresponds to the route-1 (route-2 and route-3) in Figure 2(a).

Concerning the horizontal configuration (corresponding to a vertical crystalline interface), the calculated eigenfrequencies for all the wavenumbers $k_\parallel$ parallel to the interface are highlighted in red (yellow) in Figure 3(a) for $\gamma = 1$ ($\gamma = 0.1$). The (dark) gray region herein shows the eigenfrequencies calculated under PBC. The spectral area calculated within the projected band framework almost coincides with the PBC results, indicating no significant changes exist. The line gaps denoted by the dashed black lines keep intact also, further confirming that the interface is topologically trivial. However, the spectral area alters in the oblique configuration, as shown in Figure 3(c). We see that the spectral area calculated by the projected setup (OBC) mainly lies inside that of PBC, and the density of eigenfrequencies changes compared with Figure 3(a). By checking the spectral area for each $k_{\parallel}$, we find that all the closed loops of PBC in the complex eigenvalue plane collapse in the OBC. Recalling Figure 2(d), Figure 3(c) tells that the point gaps have been formed and reveals that this interface is topologically nontrivial. Such an intrinsic difference between the eigenfrequencies under PBC and the BC-sensitive spectral patterns is a typical feature in the non-Hermitian scenario (see Section II in Supplemental Materials for comparing between the PBC and OBC results). The consistency of Figures 2 and 3 undoubtedly indicates that the two crystalline interfaces are different in nature.

Since the skin effect means that eigenmodes within a frequency range are all localized to the interface, we add the eigenmodes under consideration in an incoherent way as

$$W(\boldsymbol{r}) = \frac{1}{N_s}\sum_n w_n(\boldsymbol{r}) = \frac{1}{N_s}\sum_n[\varepsilon(\boldsymbol{r})|\boldsymbol{E}_n(\boldsymbol{r})|^2 + \mu_0|\boldsymbol{H}_n(\boldsymbol{r})|^2], \qquad (2)$$

where $w_n(\boldsymbol{r})$ with $n$ running from 1 to $N_s$ denotes the energy density distributions of the $n$-th eigenstate [52], $\boldsymbol{E}_n(\boldsymbol{r})$ and $\boldsymbol{H}_n(\boldsymbol{r})$ are the electric and magnetic field of the $n$-th eigenstate, $\varepsilon(\boldsymbol{r})$ is the dielectric constant of the calculated structure, and $\mu_0$ is the permeability of the vacuum. Such a definition smears out the phase information, and thus $W(\boldsymbol{r})$ describes the overall distribution characteristic of the considered modes. In the horizontal configuration, Figure 3(b) shows that $W(\boldsymbol{r})$ of the eigenmodes surrounded by the line gaps in Figure 3(a) distributes uniformly in the bulk region, which illustrates that all eigenmodes here are bulk modes. We then perform the same calculations to study the modes pointed by the blue arrow in Figure 3(c), and the $W(\boldsymbol{r})$ distributions are shown in Figure 3(d). These modes are clearly localized to the interface with strengths exponentially decaying as leaving away from the interface, showing the typical feature of skin modes. Although the decay rate highly depends on the non-Hermitian strength, the fields always decay in an exponential form in Figure 3(d) (see Section III in Supplemental Materials for details). The comparison between Figures 3(b) and 3(d) nicely demonstrates the GDSE, *i.e.* the crystalline interface corresponding to the route-2 and route-3 supports the skin effects, while the interface corresponding to the route-1 does not. Before proceeding, it is worthy pointing out that our system is Lorentz reciprocal, which already implies the spectral reciprocity. However, the crystalline interface breaks the spectral reciprocity from the macroscopic point of view, generating the desired GDSE (see Section IV in Supplemental Materials for the basic logic flow).

**4 Time evolution of a pulse**

Considering the excitations of GDSE from the experimental aspect, we construct two finite-sized systems exhibiting triangle and rectangle shapes, respectively, as shown in Figure 4(b). While the triangle system has two trivial interfaces and one non-trivial interface which supports the GDSE, four interfaces of the rectangle system are all trivial. To probe the GDSE, we excite a Gaussian pulse at the center of two systems and then study its time evolutions. As shown in Figure 4(a), the frequency component of the pulse covers the GDSE frequency of interest claimed in Figure 2. Since the pulse is off after $10t_0$, we show in Figure 4(b) the evolution of the pulse beginning from $20t_0$. At $t = 20t_0$, the wavefront in both structures does not see the PEC boundaries yet and possesses an elliptical form which is determined by the group velocity of the photonic bands. If nothing non-trivial happens, the waves incident on the edges will experience specular reflections. As time passes by, we see in Figure 4(b) that the waves touching the vertical edges form the standing wave patterns, showing a strong specular reflection from $t = 35t_0$ to $t = 50t_0$. On the other hand, the waves touching the horizontal edges are also specularly reflected although the wavefront is cylindrical because of the incident pulse. However, when the pulse touches the oblique edge, the waves immediately turn to the upper-left and lower-right direction along the oblique edge, stamped as yellow arrows in Figure 4(b) (see Section V in Supplemental Materials for the comparison with the Hermitian case). This hints that most of the wave components, no matter directly excited from the source or reflected by the other edges, are deflected along the edge dominantly as long as they are scattered by the interface. This indicates that the pulse is imprinted by the skin effects which are unique to the oblique edge, and hence confirms the GDSE (see movies in Supplemental Materials for details).

To investigate the power flux in the triangular configuration, we perform the continuous wave excitation calculation in the frequency domain. The system is excited by a line source placed at the white point shown in Figure 4(d), and we choose two detection points near the edges highlighted by the black and blue circles in Figure 4(d). The detection points are far from the line source to ensure that the bulk modes are not recorded. The spectra of the power flux perpendicular to the vertical and oblique edge ($S_\perp$) are respectively shown by the black and blue markers in Figure 4(c). We show the norm of $S_\perp$ in the log scale, while the sign of $S_\perp$ is dictated by the marker types, since the power flux perpendicular to the interface can be either inward or outward from the edge. Figure 4(c) tells that the $|S_\perp|$ for the oblique edge is much larger than that for the vertical edge for all the frequencies, and no outward $S_\perp$ can be found to the oblique edge, hinting negligible reflective waves. In comparison, both inward and outward $S_\perp$ are clearly seen to the vertical edge, indicating that the waves experience the specular reflection at the vertical interface. The power flux distribution at the frequency (ii) depicted in Figure 4(d) shows that the power flux bends towards the oblique edge where the GDSE exists. This is consistent with the transport behavior of skin effects in the 1D system [73, 74], further confirming the GDSE (see Section VI in Supplemental Materials for more details of the frequency-domain calculation).

Figure 4 reveals that the skin effects require the propagation of waves to probe, and thus the point gap positions relative to the real frequency axis are crucial in reality. Hence, the averaged magnitudes of $\text{Im}[\omega]$ for a point gap in consideration shall be smaller than its overall real frequency range. Otherwise, the decay process will dominate the wave behavior, which hinders the observation of GDSE. Concerning the calculation in Figure 4, the averaged $\text{Im}[\omega]$ (real frequency range) for the point gap near the EP claimed in Figures 2b and 2c are about 10 THz (100 THz), making the GDSE observation possible. In other words, the imaginary part of eigenfrequencies cannot be ignored in the transport and excitation scenarios [75].

## 5 Conclusion

To summarize, we propose a strategy for designing a 2D PhC system possessing GDSE. Via the point gaps implied from the order-2 EPs with nonzero eigenvalue winding numbers, the spectral area can be obtained along a particular wavevector direction that does not align with any high symmetry axes of the structure. The crystalline interface corresponding to this wavevector direction supports the skin effect, which has been verified by both the projected band structure calculation and the transport calculation in the time and frequency domain. Our results suggest an alternative way to realize the skin effects in the 2D (3D) system because the order-2 exceptional points (lines) can be found without any prior symmetry requirement in the non-Hermiticity. Thus, the non-Hermiticity can be either from the intrinsic losses or attained by radiative losses of a photonic crystal slab [76]. From the reciprocity theorem point of view, we effectively replace the overall wavenumbers with the wavenumber normal to the interface [68]. Such an approach, together with a myriad of tricks done in the reciprocity of metamaterials, provides a wonderland for manipulating the transport and energy partition of electromagnetic waves [77, 78]. Last and not the least, the interplay between the eigenvalue topology of EPs and the spectral topology in the non-Hermitian band structures offers a platform to manipulate the NHSEs in a sophisticated wave system when the multiple order-2 EPs and higher-order EPs are present.

**Acknowledgements:** We thank Prof. Chen Fang and Dr. Zhesen Yang for helpful discussions.

**Funding:** This work was supported by the Natural Science Foundation of China (No. 12174072) and Natural Science Foundation of Shanghai (No. 21ZR1403700).

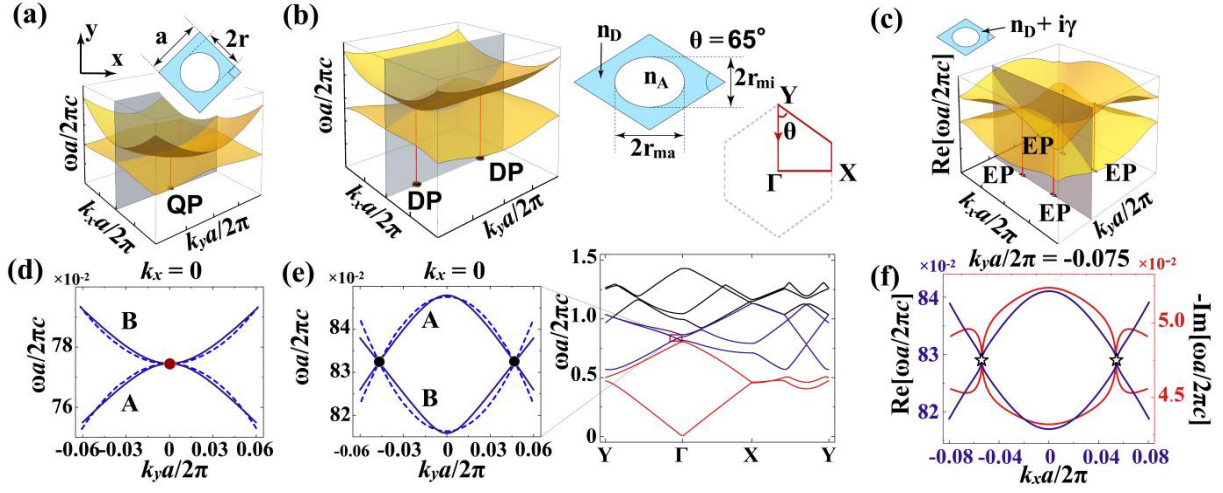

Figure 1. (a) Band structure near the Γ point for a square lattice of air holes ($n_A = 1$) in dielectric ($n_D = 1.46$). Shown in the top is the unit cell with the lattice constant $a$ and the hole radius $r$ labeled herein. (d) The bands along the $k_x = 0$ direction [the gray surface in (a)]. The filled red circles in (a) and (d) highlight the QP at the Γ point. (b) Band structure near the Γ point for an orthorhombic lattice. The rhombic unit cell depicted in the right panel is characterized by the lattice constant $a$, the angle $\theta$, and the major (minor) axis of the elliptical hole $r_{ma}$ ($r_{mi}$). (e) The full band structure along the Y − Γ − X − Y direction is shown in the right-hand side, and the bands forming the DPs are shown in an enlarged view on the left-hand side. The filled black circles in the left panel of (b) and (e) denote the DPs in the Γ − Y direction. The A and B in (d) and (e) indicate the band symmetry. (c) The Re(ω) surface near the Γ point for the orthorhombic lattice with an additional loss term in the dielectric as $n_D + i\gamma$. (f) The values of Re(ω) [Im(ω)] along the gray surface in (c) are shown by the blue [red] lines. The open black stars in (c) and (f) label the EPs spawned from the DP. The bands in (a), (b), and (c) and the dashed line in (d) and (e) are from the model Hamiltonian, while the solid line in (d), (e), and (f) are calculated by FEM for the Ez polarization. The parameters used are $a = 1365$ nm, $r = 520$ nm, $\theta = 65°$, $r_{ma} = 1.19r$, and $r_{mi} = 0.76r$. The non-Hermitian strength in (c) and (f) is $\gamma = 0.1$.

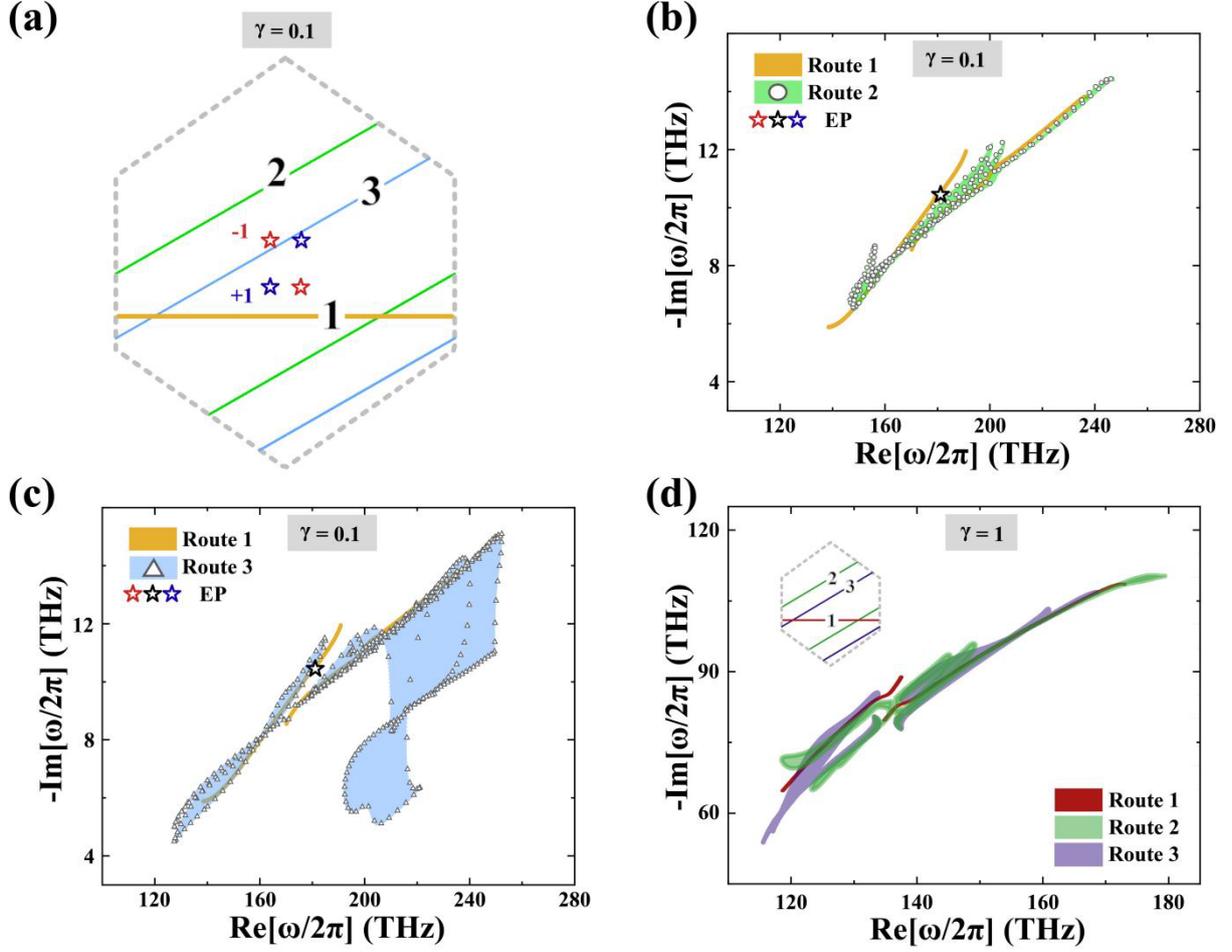

Figure 2. (a) The four order-2 EPs in the first BZ are shown by the open stars with their color displaying the sign of the DN. The solid orange, green, and blue lines show three different routes of interest. (b) The eigenfrequencies along the route-1 and route-2 are shown in the complex frequency plane by the orange lines and open circles, respectively. The profile of the spectral area spanned by the eigenfrequencies along the route-2 is highlighted in green. (c) The eigenfrequencies along the route-3 are shown by the open triangles in the complex frequency plane. The spectral area spanned by the eigenfrequencies along the route-3 is highlighted in blue. The open stars in (b) and (c) denote the EPs, and the non-Hermitian strength in (a-c) is $\gamma = 0.1$. (d) The spectral areas along the same routes with (a-c) for $\gamma = 1$. All other parameters are the same with Figure 1.

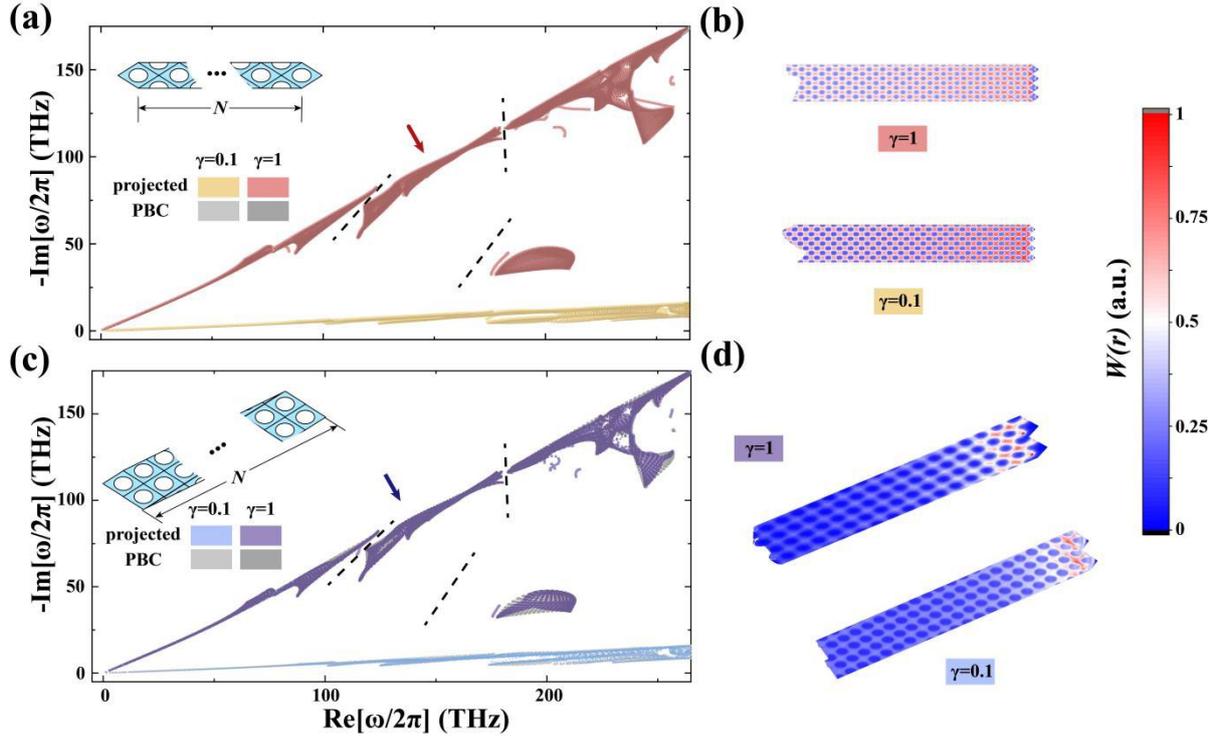

Figure 3. The projected band structure shown in the spectral area view for the crystalline interfaces perpendicular to (a) the route-1 and (c) the route-2/3. The inset in (a) and (c) schematically shows the simulation setup, and the lattice constants are $2a \sin(\theta/2)$ and $a\sqrt{5 - 4\cos(\theta)}$, respectively. The spectral area with a dark (light) color represents the $\gamma = 1$ ($\gamma = 0.1$) case. The gray color denotes the spectral area under PBC, and the black dashed lines denote the line gaps. The spatial distributions of $W(r)$ corresponding to the configuration (a) and (c) are shown in (b) and (d), respectively. The eigenstates under consideration are indicated by the red and blue arrows. The number of unit cells $N$ used in the calculation is 50.

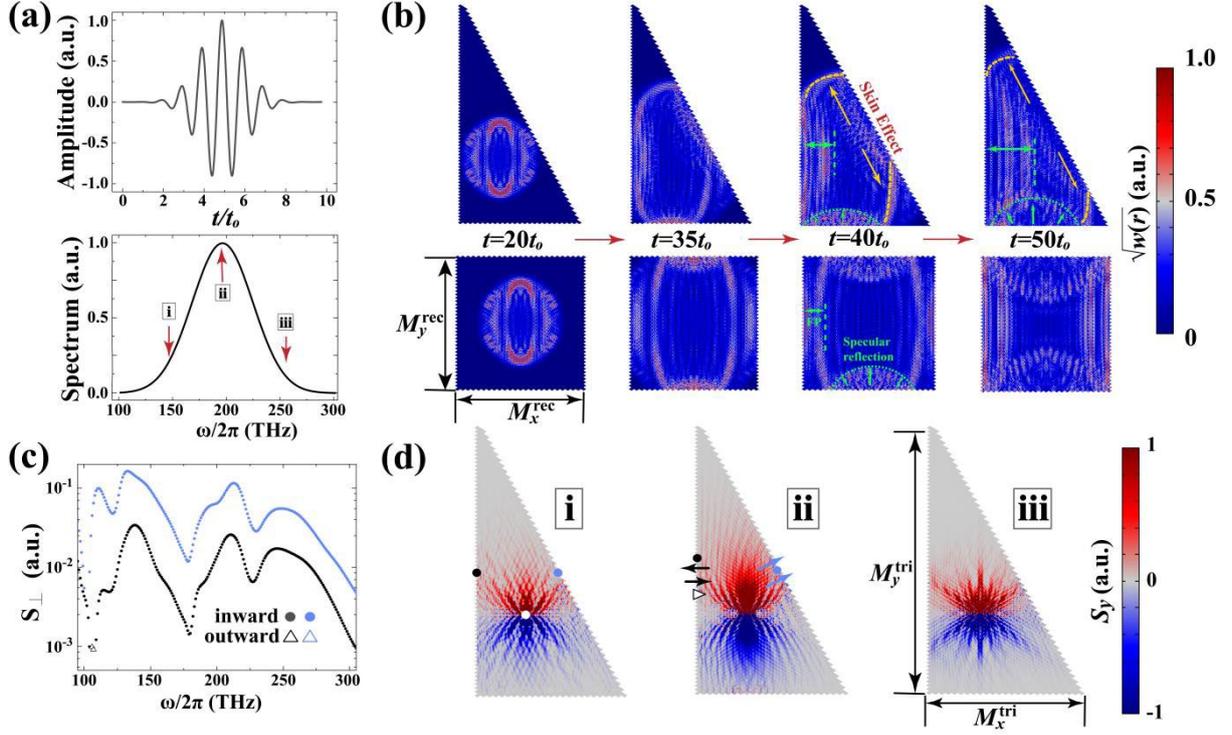

Figure 4. (a) The profile of an input Gaussian pulse in the time domain and its spectrum in the frequency domain. The Gaussian pulse is defined by $J_z = \cos(2\pi f_0(t - 1/f_1)) \times \exp\{-[\pi f_0(t - 1/f_1)]^2\}$, where $f_0 = 195$ THz and $f_1 = 40$ THz are the central frequency and standard deviation respectively. The time step used in this figure is $t_0 = 1/f_0 = 5.1$ fs. (b) The calculated time evolution of the Gaussian pulse excited inside a triangle and a rectangle. The color bar is set to square root of the energy density. The dashed green and orange lines respectively sketch out the specular reflection and the skin effects. (c) The power flux $S_\perp$ at the blue and black dots marked in (d) as a function of frequency are highlighted in blue and black. The filled circles (hollow triangles) denote that the averaged power flux is into (away from) the interface. (d) The patterns of $S_y$ for three frequencies labeled as (i-iii) in the bottom panel of (a). The parameters are $\gamma = 0.1$, $M_x^{tri} = 24$, $M_y^{tri} = 72$, $M_x^{rec} = 25$, and $M_y^{rec} = 40$.

# Supplemental Materials --- Geometry-dependent skin effects in reciprocal photonic crystals

Zhening Fang, Mengying Hu, Lei Zhou*, and Kun Ding†

Department of Physics, State Key Laboratory of Surface Physics, and Key Laboratory of Micro and Nano Photonic Structures (Ministry of Education), Fudan University, Shanghai 200438, China

*Email: phzhou@fudan.edu.cn

†Email: kunding@fudan.edu.cn

## Section I. A two-level Hamiltonian model for the quadratic point and exceptional points

The two-level Hamiltonian which can describe the quadratic point (QP) in the square lattice is read as [S1]

$$H_0 = (\omega_Q + \mu k^2)\boldsymbol{\sigma}_0 + \beta(k_x^2 - k_y^2)\boldsymbol{\sigma}_3 + 2\lambda k_x k_y \boldsymbol{\sigma}_1, \qquad (S1)$$

where $\omega_Q$ is the frequency of the QP, $\boldsymbol{\sigma}_i$ ($i = 0, 1, 2, 3$) are Pauli matrices, and $k^2 = k_x^2 + k_y^2$. The values of $\beta$, $\lambda$, and $\mu$ determine the dispersions of the bands along different directions. Note that the above Hamiltonian works near the Brillouin zone (BZ) center, and we are interested in the bands along the $\Gamma - Y$ direction. Through fitting the finite element method (FEM) results, the parameters used in Figure 1(d) of the main text are $\omega_Q = 2\pi \times 170.2$ THz, $\beta = 1200 \times \left(\frac{a}{2\pi}\right)^2$ THz, $\mu = -100 \times \left(\frac{a}{2\pi}\right)^2$ THz, and $\lambda = 0$.

Since the Dirac points (DPs) are attained by deforming the square lattice, the Hamiltonian shall have the symmetry breaking terms, which can be written as

$$H_1 = H_0 + H_b = H_0 + \sum_{i=1,2,3} \alpha_i \boldsymbol{\sigma}_i, \qquad (S2)$$

where $\alpha_i$ are the parameters controlling the different ways of breaking the $C_{4v}$ symmetry. $\alpha_1$ corresponds to changing the relative length of the two lattice vectors, $\alpha_2$ corresponds to preserving the $C_4$ rotational symmetry but breaking reflection symmetries, and $\alpha_3$ corresponds to shearing the lattice which preserves the reflection symmetry with respect to the $x$ and $y$ axes but breaks the $C_4$ rotational symmetries. For the structure under consideration as shown in Figure S1(a), only $\alpha_3$ is allowed to be nonzero, and the resulting eigenvalues are

$$\omega_\pm = \omega_Q + \mu k^2 \pm \sqrt{(2\lambda k_x k_y)^2 + [\alpha_3 + \beta(k_x^2 - k_y^2)]^2}. \tag{S3}$$

To fit the bands shown in Figure 1(e) of the main text, the parameters are $\omega_Q = 2\pi \times 183.2$ THz, $\beta = 1200 \times \left(\frac{a}{2\pi}\right)^2$ THz, $\mu = -100 \times \left(\frac{a}{2\pi}\right)^2$ THz, $\lambda = 300 \times \left(\frac{a}{2\pi}\right)^2$ THz, and $\alpha_3 = 6.8 \times \left(\frac{a}{2\pi}\right)^2$ THz.

The Hamiltonian forms for the non-Hermitian system are the same as Eq. (S3), but the parameters herein shall be complex numbers in general. As discussed in the main text, if one parameter or one particular set of parameters become complex and satisfy the pseudo-Hermitian condition, then the DP can give rise to an exceptional ring. However, if all the parameters become complex, then only a pair of EPs can be found near the DP, which is the case shown in Figure 1 of the main text. To describe the behavior near EPs clearly, we set the parameters to $\omega_Q = 2\pi \times (183.2 + 11i)$ THz, $\beta = (1200 + 500i) \times \left(\frac{a}{2\pi}\right)^2$ THz, $\mu = -(100 + 20i) \times \left(\frac{a}{2\pi}\right)^2$ THz, $\lambda = 300 \times \left(\frac{a}{2\pi}\right)^2$ THz, and $\alpha_3 = (6.8 + 10i) \times \left(\frac{a}{2\pi}\right)^2$ THz. The real part of the modeled Hamiltonian is shown in Figure 1(c) of the main text, while the image part is shown in Figure S1(b).

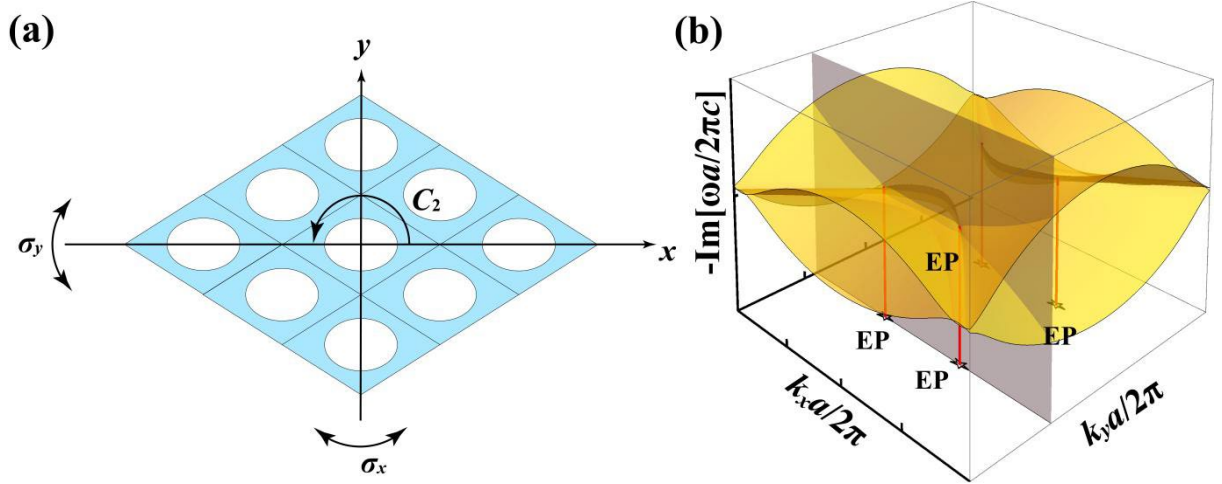

**Figure S1.** (a) The symmetry operations for the calculated system. (b) The image part of the modeled Hamiltonian. The grey surface is the same with that in Figure 1(c) of the main text located at $k_y a/2\pi = -0.123$.

## Section II. The spectral patterns under periodic and open boundary conditions

As we have shown in the main text, there shall be infinite parallel routes in the reciprocal lattice space perpendicular to the same crystalline interface in the real space. To demonstrate how to choose these routes, we depict in Figure S2 the route-1, route-2, and route-3 under consideration in the extended zone scheme. All routes can form a close loop in the BZ. As shown in Figure S2, the location of the $\Gamma$ points are: $(0,0)$ for $\Gamma^{(1)}$, and $\frac{\pi}{\cos(\theta/2)a}\hat{x} + \frac{\pi}{\sin(\theta/2)a}\hat{y}$, $\frac{\pi}{\cos(\theta/2)a}\hat{x} - \frac{\pi}{\sin(\theta/2)a}\hat{y}$, $-\frac{\pi}{\cos(\theta/2)a}\hat{x} + \frac{\pi}{\sin(\theta/2)a}\hat{y}$, $-\frac{\pi}{\cos(\theta/2)a}\hat{x} - \frac{\pi}{\sin(\theta/2)a}\hat{y}$, $\frac{2\pi}{\cos(\theta/2)a}\hat{x}$, $-\frac{2\pi}{\cos(\theta/2)a}\hat{x}$ for $\Gamma^{(2)}$, respectively. The criteria for route-2 and route-3 forming the closed $k$ loops in the BZ are $\overrightarrow{PQ}||\overrightarrow{P'Q'}$, $\overrightarrow{PP'} = \vec{b}_1$, and $\overrightarrow{QQ'} = \vec{b}_1 + \vec{b}_2$. By solving the equations, we find that the slope of the route $\overrightarrow{PQ}$ must be fixed while the intercept can be arbitrary.

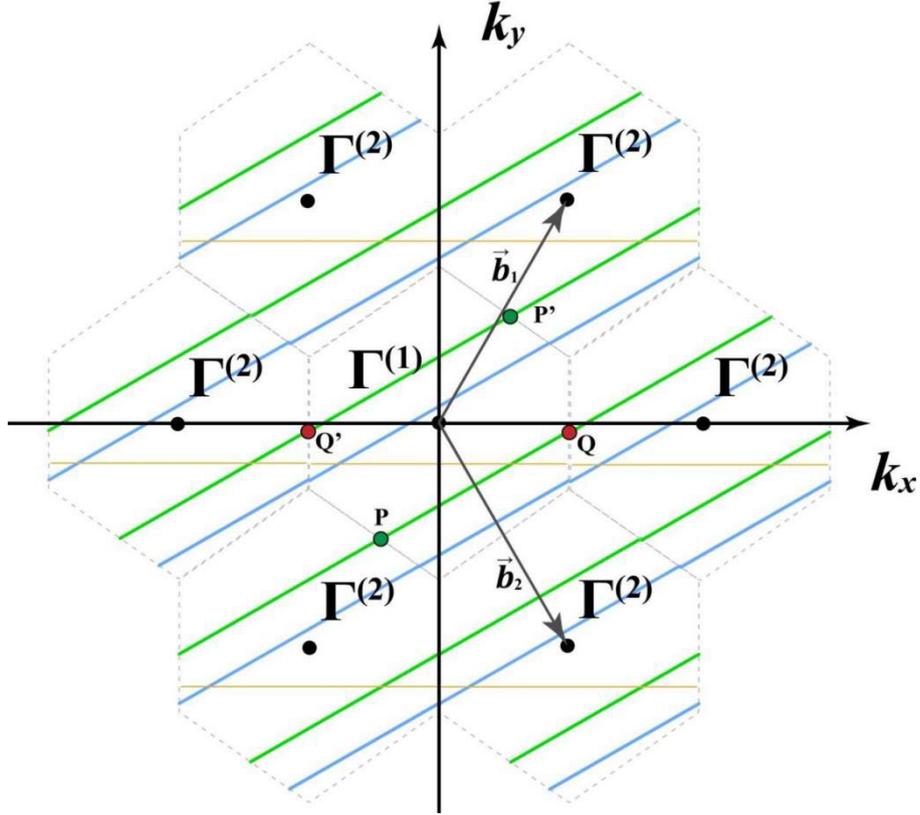

**Figure S2.** Reciprocal lattice space in the extended zone scheme. Since the point P (Q) and P' (Q') are equivalent, the segment P-Q-Q'-P' (the green route) forms a close loop in the $k$-space, and the same goes to the blue line. The reciprocal lattice vectors are $\vec{b}_1 = \frac{\pi}{\cos(\theta/2)a}\hat{x} + \frac{\pi}{\sin(\theta/2)a}\hat{y}$ and $\vec{b}_2 = \frac{\pi}{\cos(\theta/2)a}\hat{x} - \frac{\pi}{\sin(\theta/2)a}\hat{y}$.

Since we only show the spectral winding under the PBC in Figure 2 of the main text, we provide more evidence in this section. The OBC results for the oblique configuration with $\gamma = 0.1$ are shown with the PBC bands in Figure S3. As is indicated, the eigenvalues of OBC lying inside the spectral area of PBC belong to the corresponding bands, while the other eigenvalues belong to other energy bands of PBC. The PBC bands with a spectral area or a point gap collapse to a line in the OBC calculations, which precisely demonstrates the mechanism of NHSE.

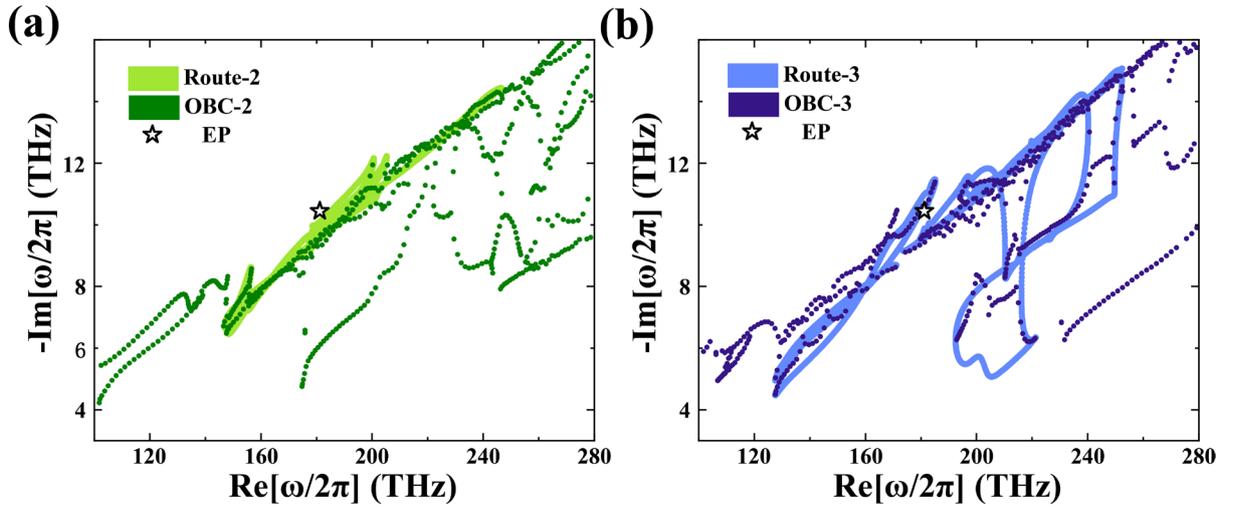

**Figure S3.** The comparison between spectral patterns calculated under the PBC [same as Figure 2(b-c) in the main text] and the OBC results (darker green and blue dots) for the (a) route-2 and (b) route-3. The black stars denote the EP. The parallel *k*-vector for OBC-2 (OBC-3) is $k_{\parallel,2} = 0.24 \times \frac{2\pi}{a}$ ($k_{\parallel,3} = 0.06 \times \frac{2\pi}{a}$), which corresponds to route-2 (route-3).

For a holistic view of the bands, the spectral windings for the lowest 8 bands when $\gamma = 1$ are shown in Figure S4. We also calculate all the routes in the reciprocal lattice space with the same slope, finding that once one of the routes can form the spectral area, all the others also can, except the high symmetry axes. Note that only the 1st lowest band does not show a spectral area along the oblique interface, while none of the bands shows the spectral area for the horizontal configuration.

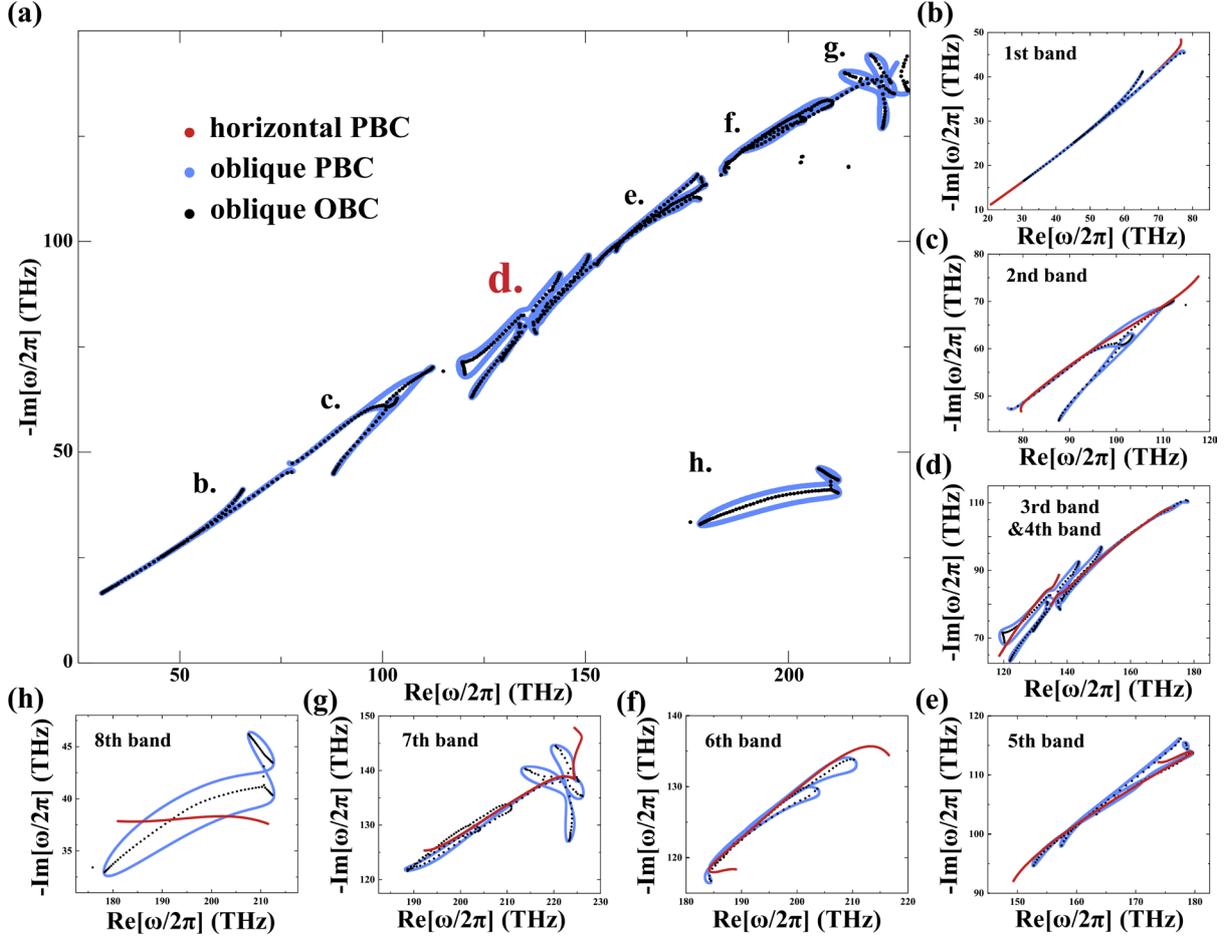

**Figure S4.** The comparison between spectral patterns calculated under the PBC (blue dots) and OBC (black dots) for the oblique configuration claimed in the main text. The panel (a) shows a full view of the 8 bands, and the enlarged views are shown from panel (b) to (h). The PBC results for the horizontal configuration is also shown by the brown dots in (b-h). Note that (b-c) the lowest 2 bands and (e-h) the 5$^{th}$ to 8$^{th}$ bands are not focused in the work, and (d) the bands containing the order-2 EPs are mostly discussed in the main text. For the OBC results, the PBC is still applied along the crystalline interfaces, and without losses of generality we choose $k_{\parallel} = 0.4 \times \frac{2\pi}{a \sin(\theta)}$. The trajectory of route-1 is $k_y = -0.15G$, and the trajectories of the route-2 and route-3 are $k_y = -0.52k_x + 0.28G$ and $k_y = -0.52k_x - 0.34G$ respectively. All other parameters are the same with Figure 3 in the main text.

As pointed out in the main text, the key to forming non-trivial ***k*** loops is the mismatch between symmetries, which means various slopes can be chosen. Figure S5(a) shows another loop in BZ, where the points $V$ and $V'$ are equivalent in the first BZ. The criterion for the specific route is simply derived as: $\overrightarrow{VV'} = \vec{b}_1$. As shown in Figure S5(b), the spectral area in the PBC calculation persists and also collapses to several lines under the OBC. Therefore, the skin effects can still be observed for different non-Hermiticity strengths, as shown in Figure

S5(c). We note here that such a method to obtain the non-trivial closed ***k*** loops is only for the convenience of constructing the crystalline interfaces composed of perfect unit cells. However, the existence of skin effects does not require the chosen ***k*** loops to close solely in the first or extended BZ.

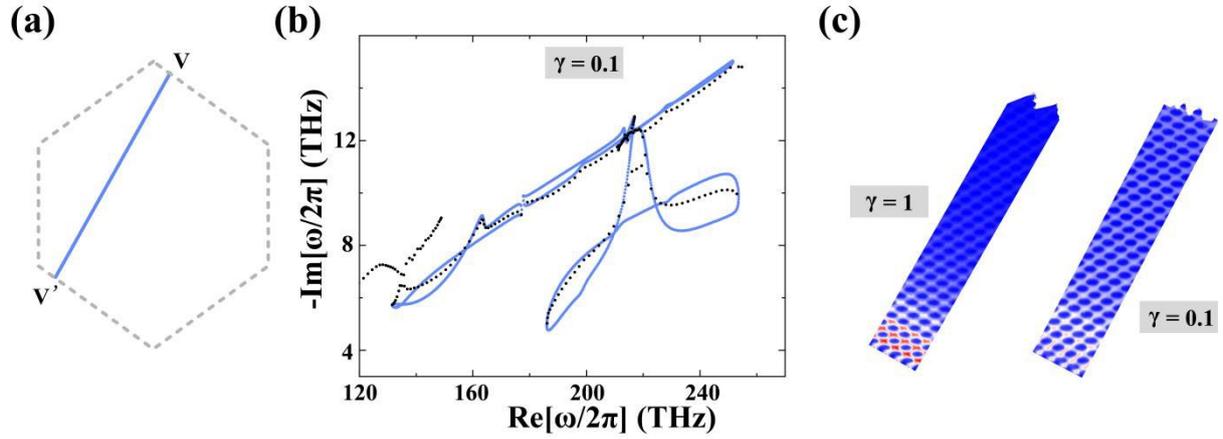

**Figure S5.** (a) Another chosen route in the BZ where the segment V-V' (the blue route) forms a close loop in the *k*-space. (b) The spectral areas formed by the chosen route in the PBC and the collapsed bands in the OBC are respectively shown by the blue lines and black dots. (c) The skin effects formed by the bands in (b) with two different non-Hermitian strengths are compared. The parallel *k*-vector in the OBC is $k_{||} = 0.25 \times \frac{2\pi}{a}$, corresponding to the blue route in (a).

**Section III. Numerical verification of geometry dependent skin effects**

    We have demonstrated in the main text the GDSE based on the one-dimensional setup using in the projected band structure calculations, and the number of unit cells $N$ is fixed to 50. In this section, we increase $N = 100$ to further verify the GDSE, as shown in Figure S6. The distributions of $W(x)$ in Figures S6(a) and S6(b) indicate whatever the values of $\gamma$, the skin effects always exist in the oblique configuration, but do not in the horizontal configuration. This reveals the geometry-dependent nature, and we shall further verify the asymptotic decay behavior of $W(x)$ near the interface to confirm the GDSE. Figures S6(c) and 6(e) show $W(x)$ along the black dashed line highlighted in Figures S6(a) and S6(b) for two different values of $\gamma$. Besides the geometry-dependent nature, it is also clear that the modes are more likely to localize when $\gamma$ is larger. To quantify, we use a semi-log plot to $W(x)$ near the interface, as shown in Figures 6(d) and 6(f). The linear dependence is clearly seen, indicating an exponential decay of $W(x)$. The decay rate is dominated by the values of $\gamma$, which reveals the modes are unique to the crystalline interface. This finishes numerical verification of the GDSE.

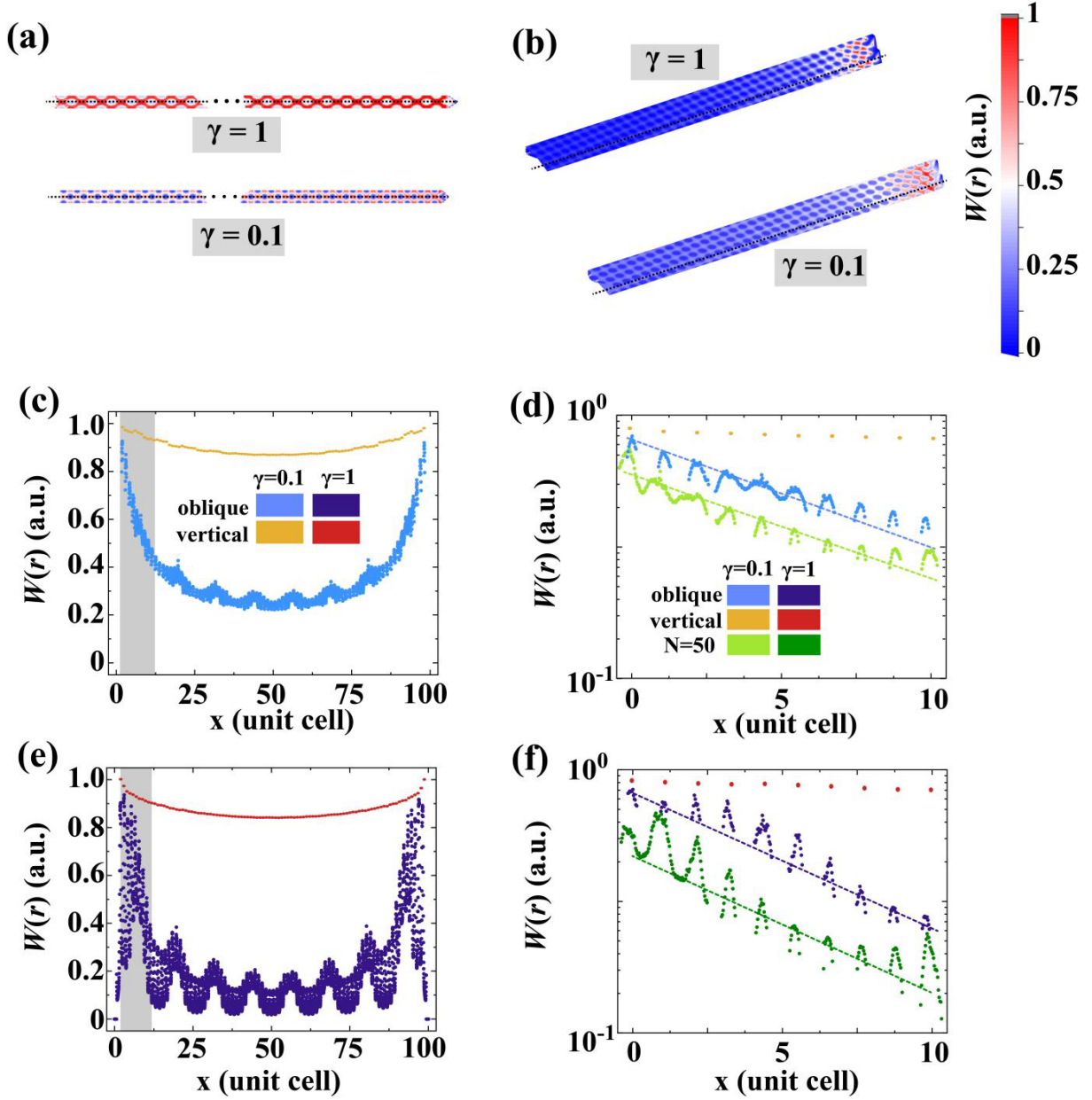

**Figure S6.** The two-dimensional distribution of $W(x)$ for the (a) horizontal and (b) oblique configuration when $N = 100$. The values of $W(x)$ as a function of the position away from the interfaces for (c) $\gamma = 0.1$ and (e) $\gamma = 1$ at different interface positions. The semi-log plot of $W(x)$ near the interface shown as the grey area for the results in (c) and (e) is shown in (d) and (f), respectively. The blue and purple (orange and red) dots in (c-f) stand for the oblique (horizontal) configuration, while the green and olive dots in (d) and (f) stands for the oblique configuration when $N = 50$, which is the same as the main text. The dashed line shows that the energy density follows the exponential decay from the first unit cell. The fitted curves are $w_{\gamma=0.1}(x) = \exp\{-0.05x\}$ for $\gamma = 0.1$ and $w_{\gamma=1}(x) = \exp\{-0.07x\}$ for $\gamma = 1$.

## Section IV. The flowchart of our design strategy

As shown in Figure 2 of the main text, a stable EP in the 2D PhC is guaranteed to have a nonzero DN, while the EP is not a must for the DN. Once the nonzero DN is observed, the symmetry of the unit cell and the overall structure are compared. By mismatching the two symmetries, a closed route in the $k$-space is chosen to ensure the spectral area in the complex frequency plane. With the standing wave argument in the main text, such a spectral area is equivalent to the existence of skin effects upon the corresponding interface.

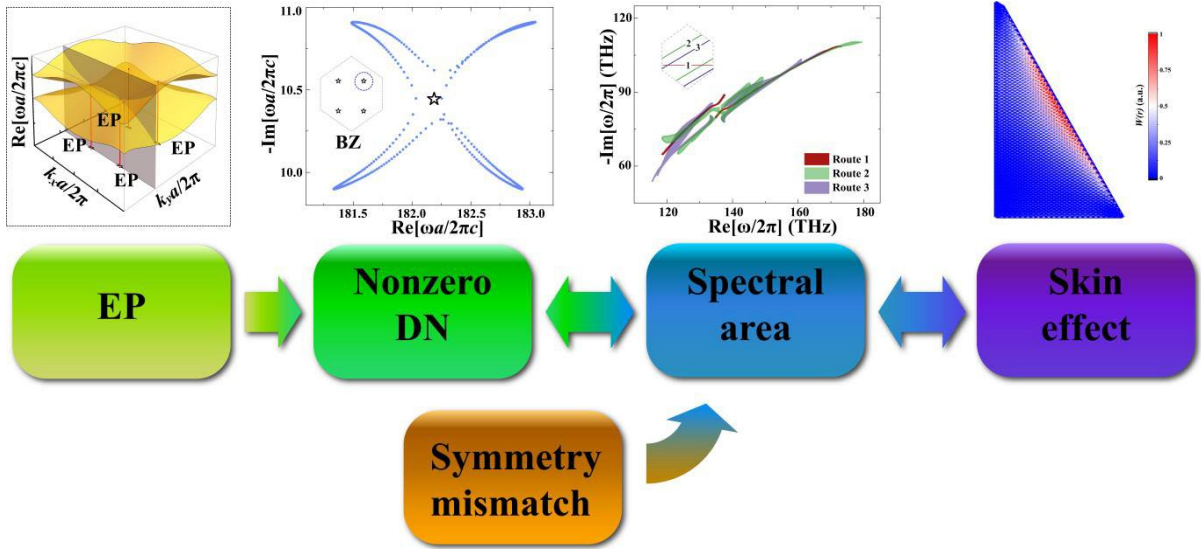

**Figure S7.** The flowchart of our design. The double arrows denote the equivalency, while the single arrow herein means a sufficient but not necessary condition.

## Section V. Comparison with the Hermitian scenario

We have shown the evolution of a pulse inside a triangle and a square for the same value of $\gamma$ in Figure 4 of the main text to demonstrate the non-trivial crystalline interface. Here, we compare the triangular configuration possessing non-trivial skin effects ($\gamma = 0.1$) with the same configuration without the skin effects by setting $\gamma = 0$. Figure 4(c) of the main text has been duplicated as Figure S8(a), and Figure S8(b) shows the results when $\gamma = 0$. As shown in Figure S8(b), the waves touching the oblique edges show the spectral reflections, while the spectral reflections are not seen in Figure S8(a). This also illustrates that skin effects are found in our non-Hermitian system.

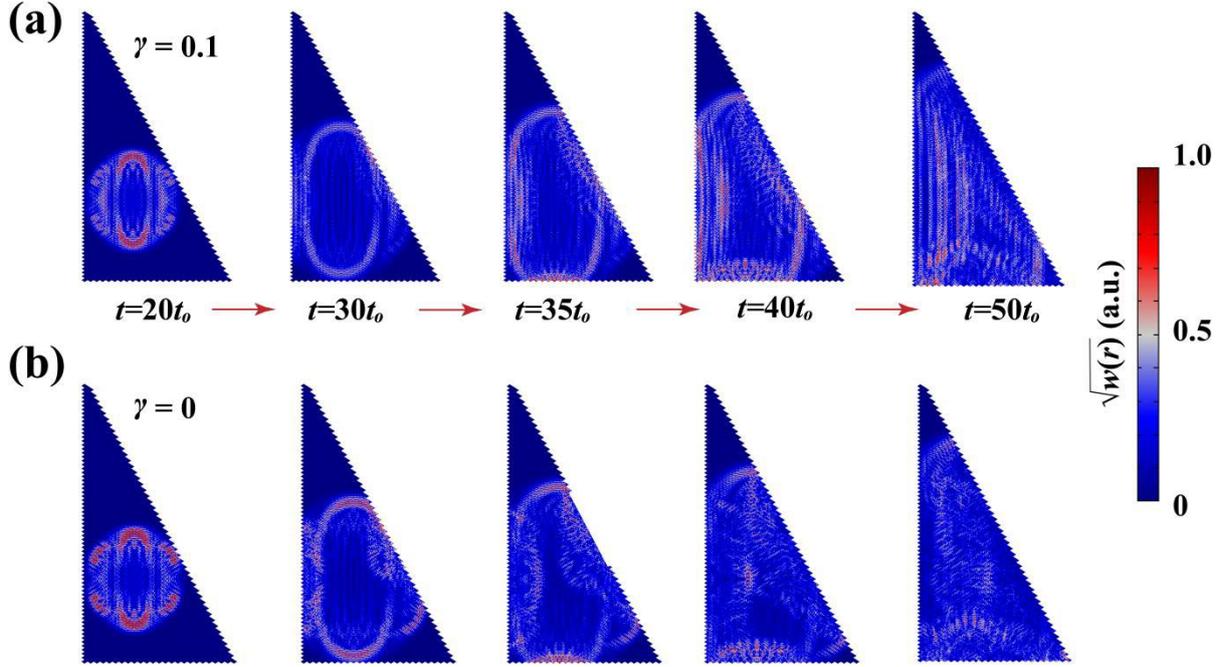

**Figure S8.** The dynamic evolution of a Gaussian pulse inside a 2D triangle for (a) $\gamma = 0.1$ and (b) $\gamma = 0$. Strong specular reflection is seen for the oblique edge of $\gamma = 0$, with a clear wavefront at $t = 30t_0$ and $t = 35t_0$, while strong interference with the reflection field of the vertical edge is shown at $t = 40t_0$ and $t = 50t_0$. All the parameters are the same with Figure 4 in the main text.

**Section VI. Detailed analysis of the power flux in the triangular configuration**

To analyze the power flux shown in Figure 4(c) of the main text more fairly, we move the source to the point marked by O in Figure S9(a) and record the waves at p, q, s, and t. To quantify the locations, we denote the unit cell in the bottom left of the triangle as (1,1), and then any unit cell can be denoted as ($m$, $n$). The first (second) letter in the parenthesis denotes the distance of the unit cell with respect to (1,1) in the horizontal (vertical) direction. With such notation, the location of the excited spot O is (5, 6), while the position of p, q, s, and t are (1, 37), (9, 37), (16.5, 10.5), and (16.5, 1.5), respectively. Since the edge of the realistic structure is formed in a zigzag way, we calculate the average power flux for the minimum period, namely four unit cells near the spots, as shown in the left and top right panel of Figure S9(a) for the spot p and q. Note that the spots p and q are symmetric about the spot O in the horizontal direction, while the spots s and t are symmetric about the spot O in the vertical direction. Therefore, the spots p and q (s and t) are treated as one pair of detecting points. We deliberately choose these points to make the comparison fair because the distances between the points in each pair and the source O are the same. Since the unit cell possesses mirror symmetry about the *x* and *y* axes, the optical length from the source to the detecting points in each pair is also the same, and the ensuing radiation power directly from the source shall be the same.

Figures S9(b) and S9(c) respectively show the total power flux perpendicular to the interface $S_\perp$ at p/q and s/t. Besides the directly radiated waves from the source, the total waves recorded at the detecting spots also have contributions from the potential skin modes and the reflected waves from the interface. However, the features of $S_\perp$ from these two contributions are distinct. Since the skin effect occurs when the incident wave is not reflected specularly, $S_\perp$ shall be always towards the edge, but for a conventional interface with reflective waves, $S_\perp$ can be either towards or away from the edge. As a result, examining $S_\perp$ near the interface offers a simple way to identify the skin effect. The comparison between the blue and black markers in Figures S9(b) and S9(c) nicely shows the above feature and thus confirms the observation of GDSE.

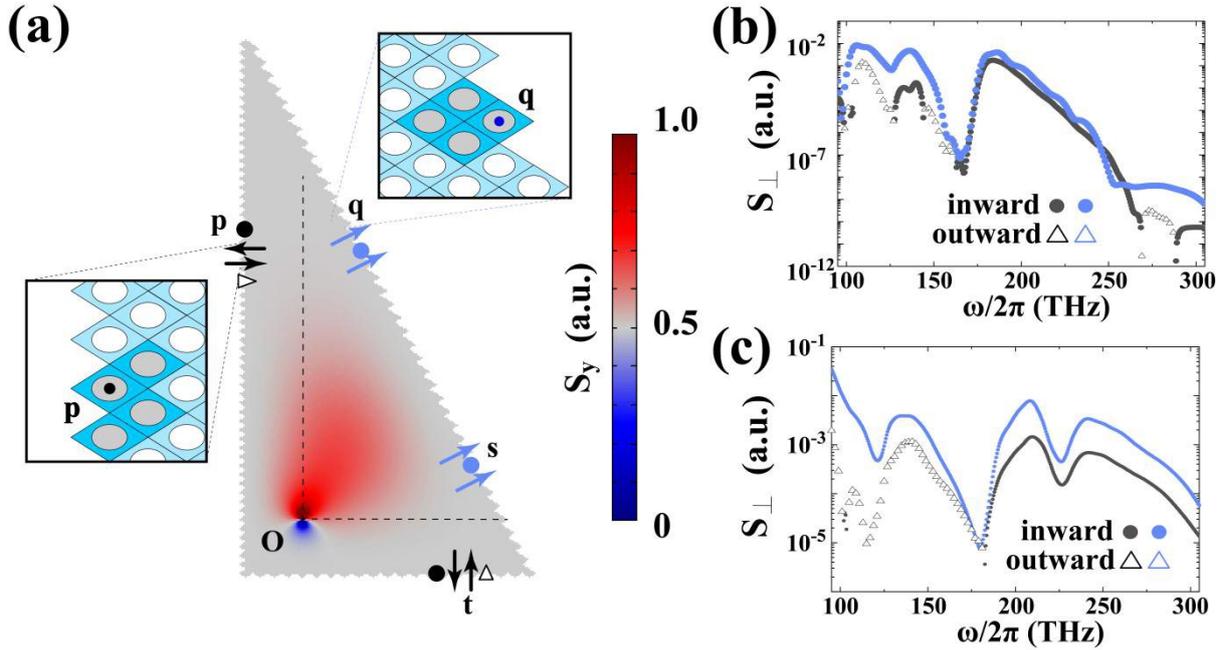

**Figure S9.** (a) A scheme of the calculation setup with the color plot showing the power flux along the y axis. The spot O indicates the position of the incident source, while p, q, s, and t are the detecting spots. (b) The calculated power flux perpendicular to the edges $S_\perp$ at the p (black) and q (blue). (c) The calculated power flux perpendicular to the edges $S_\perp$ at the s (blue) and t (black). The dots (triangles) indicate that the power flux is towards (away from) the edge.